\documentclass[12pt]{iopart}
\newtheorem{thm}{Theorem}
\newtheorem{lem}{Lemma}

\usepackage{iopams}
\begin{document}

\title[On some class of homogeneous polynomials]{On some class of homogeneous polynomials and explicit form of integrable hierarchies of differential-difference equations}

\author{A K Svinin}

\address{Institute for System
Dynamics and Control Theory, Siberian Branch of
Russian Academy of Sciences, Russia}
\ead{svinin@icc.ru}
\begin{abstract}
We introduce two classes of homogeneous polynomials and show their role in constructing of integrable hierarchies for some integrable lattices.

\end{abstract}

\vspace{2pc}


\submitto{J. Phys. A: Math. Theor.}

\section{Introduction}
Our main goal in this paper is to introduce two classes of homogeneous polynomials $T^l_s$ and $S^l_s$ of many variables and to show its applicability
in the theory of integrable differential-difference equations (lattices). More exactly, we construct in terms of these polynomials explicit form of some integrable hierarchies. We base our studies on the relationship of integrable lattices with KP hierarchy. To this aim, we consider bi-infinite sequences of KP hierarchies in the form of local differential-difference conservation laws \cite{Wilson} expressed in terms of generating relations. Our approach, in particular, goes back to \cite{Casati}, \cite{Magri}, \cite{Wilson}.

Let us give a sketch of our approach for investigation of some integrable lattices and its hierarchies. A presentation of the sequence of KP hierarchy in the form of two generating equations
\[
\fl
\partial_s h(i, z)=\partial H^{(s)}(i, z)\;\;\mbox{and}\;\; \partial_sa(i, z)=a(i, z)\left(H^{(s)}(i+1, z)-H^{(s)}(i, z)\right),\;\;i\in\mathbb{Z},
\]
where $\partial_s$ stands for a derivative with respect to the evolution parameter $t_s$, can be found in \cite{Magri}. Taking this representation as a basis  we have shown in \cite{Svinin1}, \cite{Svinin2}, \cite{Svinin3} that many integrable lattices can be obtained as reductions of this system. Our study in these papers was based on two theorems which select an infinite number of invariant submanifolds $\mathcal{S}^n_{l-1}$. In particular the restriction of this sequence of KP hierarchies on $\mathcal{S}^n_0$ gives the totality of evolution differential-difference equations on infinite number of fields $\{a_k=a_k(i)\}$. We call this set of equations the $n$th discrete KP hierarchy. We denote by $\mathcal{M}$ the corresponding phase space. Further we consider intersections of the form ${\cal S}_0^n\cap{\cal S}_{l-1}^p$ and show that this corresponds to restriction of  the $n$th discrete KP hierarchy on some submanifold $\mathcal{M}_{n, p, l}\subset\mathcal{M}$ restricting the latter to some differential-difference system of equations on finite number of fields. The restriction on  $\mathcal{M}_{n, p, l}$ is given by an infinite set of algebraic equations $J_k^{(n, p, l)}[a_1,\ldots, a_{k+l}]=0$ for $k\geq 1$.

In this paper we consider only a class of one-field lattices corresponding to submanifolds $\mathcal{M}_{n, p, 1}$ with $n\geq 1$ and $p>n$. The particular case, when $p=n+1$, is given by Itoh-Narita-Bogoyavlenskii (INB) lattice \cite{Itoh}, \cite{Narita}, \cite{Bogoyavlenskii} or extended Volterra equation in terminology of \cite{Narita}.  For general case of $\mathcal{M}_{n, p, 1}$ we construct corresponding  integrable hierarchy in explicit form. In \cite{Svinin4} we already have shown explicit form of integrable hierarchy of INB lattice equations in terms of homogeneous polynomials $S^l_s$. In this paper we construct more general class of such polynomials and corresponding integrable hierarchies.

The rest of the paper is organized as follows. In the next two sections we give explanation of the relationship of the KP hierarchy with some integrable lattices. In section \ref{sec:4}, we define two classes of homogeneous polynomials $T^l_s$ and $S^l_s$ of many variables and show some identities which directly follow from their definition. In  section \ref{sec:5}, we consider the restriction of $n$th discrete KP hierarchy on submanifolds $\mathcal{M}_{n, p, 1}$ corresponding to one-component lattices and prove there some technical lemmas. In particular, we write down an infinite number of linear equations on KP wavefunction $\Psi=\{\psi_i\}$ which corresponds to the sequence of invariant submanifolds inclusions. We show explicit form of evolution equations of some class of integrable hierarchies.   In subsections \ref{sec:7} and \ref{sec:8} we provide the reader by examples and show explicit form of some integrable hierarchies  of one-component lattices.

\section{The KP hierarchy} \label{sec:1}

\subsection{The KP hierarchy}

It is known that there exists close relationship of the KP hierarchy with some integrable lattices and its hierarchies. In this and next sections we briefly describe this relationship based on our approach developed in \cite{Svinin1}, \cite{Svinin2}, \cite{Svinin3} considering free bi-infinite chain of KP hierarchies and its suitable reductions.

We write evolution equations of the KP hierarchy itself in the form of local conservation laws \cite{Wilson}
\begin{equation}
\partial_s h(z)=\partial H^{(s)}(z),
\label{KP}
\end{equation}
where formal Laurent series $h(z)=z+\sum_{j\geq 2}h_jz^{-j+1}$ and $H^{(s)}(z)=z^s+\sum_{j\geq 1}H_j^sz^{-j}$ are related with formal KP wavefunction as $h(z)=\partial \psi/\psi$ and $H^{(s)}(z)=\partial_s \psi/\psi$, respectively. Each coefficients $H_j^s$, in fact, is calculated to be some differential polynomial $H_k^s=H_k^s[h_2,\ldots, h_{s+k}]$. More exactly, the Laurent series $H^{(s)}(z)$ is calculated as projection of $z^s$ onto the linear space ${\cal H}_{+}=<1, h, h^{(2)},\ldots>$ spanned by Fa\`a di Bruno iterates $h^{(k)}\equiv (\partial+h)^{k}(1)$. It can be shown that dynamics of coefficients $H_j^s$ in virtue of KP flows (\ref{KP}) is given by the invariance relation $(\partial_s+H^{(s)})\mathcal{H}_{+}\subset\mathcal{H}_{+}$ which can be written explicitly as
\begin{equation}
\partial_sH^{(k)}=H^{(k+s)}-H^{(k)}H^{(s)}+\sum_{j=1}^sH^k_jH^{(s-j)}+\sum_{j=1}^sH^s_jH^{(k-j)}.
\label{CS}
\end{equation}
Conversely, one can start from these equations which constitute an infinite number of commuting flows including the first flow corresponding to evolution parameter $t_1=x$ on the phase space whose points are parameterized by the semi-infinite matrix $(H_j^s)$. System (\ref{CS}) known as the Central System is equivalent in fact to the KP hierarchy and the latter can be obtained by choosing $t_1$ as the spatial variable \cite{Casati}.

\subsection{The system describing infinite chain of KP hierarchies}

Let us consider bi-infinite sequence of KP hierarchies $\{h(i, z) : i\in\mathbb{Z}\}$. To our aim, it is convenient to introduce another
Laurent series $a(i, z)=z+\sum_{j\geq 1}a_j(i)z^{-j+1}\equiv z\psi_{i+1}/\psi_i$, which as can be checked to satisfy the equation
\begin{equation}
\partial_sa(i, z)=a(i, z)\left(H^{(s)}(i+1, z)-H^{(s)}(i, z)\right).
\label{KP1}
\end{equation}
The latter in turn can be rewritten in the form looking as differential-difference conservation laws
\[
\partial_s\xi(i, z)=H^{(s)}(i+1, z)-H^{(s)}(i, z)
\]
with
\[
\fl
\xi(i, z)=\ln a(i, z)=\ln z-\sum_{j\geq 1}\frac{1}{j}\left(-\sum_{k\geq 1}a_k(i)z^{-k}\right)^j
\equiv\ln z+\sum_{j\geq 1}\xi_j(i)z^{-j}.
\]
Thus as a starting point we consider following an infinite set of evolution equations:
\begin{equation}
\partial_sh_k(i)=\partial H^s_{k-1}(i),\;\;\;
\partial_s\xi_k(i)=H^s_k(i+1)-H^s_k(i)
\label{eq3.1}
\end{equation}
which as is shown below to admit a broad class of reductions yielding an infinite number of differential-difference equations. Let us remember that the KP hierarchy is equivalent to the Central System (\ref{CS}). Therefore, we can assume that the point of our phase space is defined by infinite number of functions $\{H^j_s(i), a_k(i)\}$.

\section{The $n$th discrete KP hierarchy} \label{sec:3}

\subsection{Reductions of equations (\ref{eq3.1})}

We are going to show in this and next sections how some integrable lattices can be obtained as a result of special reductions of equations (\ref{eq3.1}). To this aim, we need in following two theorems.
\begin{thm} \label{th:1}\cite{Svinin1}
The submanifold $\mathcal{S}_{l-1}^n$ defined by the condition
\begin{equation}
z^{l-n}a^{[n]}(i, z)\in{\cal H}_{+}(i),\;\; \forall i\in{\mathbb Z}
\label{eq3.2}
\end{equation}
is invariant with respect to flows given by (\ref{eq3.1}).
\end{thm}
\begin{thm} \label{th:2}\cite{Svinin2}
The following chain of invariant submanifolds inclusions
\begin{equation}
\mathcal{S}_{l-1}^n\subset \mathcal{S}_{2l-1}^{2n}\subset\mathcal{S}_{3l-1}^{3n}\subset\cdots
\label{eq3.4}
\end{equation}
is valid.
\end{thm}

Let us spend some lines to clarify certain details. In the theorem \ref{th:1}, by definition,
\[
a^{[r]}(i, z)=z^r\frac{\psi_{i+r}}{\psi_i}=\left\{
\begin{array}{l}
\prod_{j=1}^ra(i+j-1, z),\;\;\;  r\geq 1, \\
1,\;\;\;     r=0,                            \\
\prod_{j=1}^{|r|}a^{-1}(i-j, z),\;\;\;   r\leq -1.
\end{array}
\right.
\]
Thus, the coefficients of the Laurent series $a^{[r]}(i, z)=z^r+\sum_{j\geq 1}a_j^{[r]}(i)z^{-j+r}$ are some quasi-homogeneous polynomials $a_j^{[r]}[a_1,\ldots, a_j]$. In what follows we use obvious identity
\[
a^{[r_1+r_2]}(i)=a^{[r_1]}(i)a^{[r_2]}(i+r_1)=a^{[r_2]}(i)a^{[r_1]}(i+r_2)
\]
for any $r_1, r_2\in\mathbb{Z}$, which yields an infinite set of identities
\begin{eqnarray}
a_k^{[r_1+r_2]}(i)&=&a_k^{[r_1]}(i)+\sum_{j=1}^{k-1}a_j^{[r_1]}(i)a_{k-j}^{[r_2]}(i+r_1)+a_k^{[r_2]}(i+r_1) \nonumber \\
                  &=&a_k^{[r_2]}(i)+\sum_{j=1}^{k-1}a_j^{[r_2]}(i)a_{k-j}^{[r_1]}(i+r_2)+a_k^{[r_1]}(i+r_2). \label{r1r2}
\end{eqnarray}
It is worth remarking that theorem \ref{th:1} was proven in \cite{Magri} in the case $n=1$.

It is useful to define another set of quasi-homogeneous polynomials $\{q_j^{(n, r)}=q_j^{(n, r)}[a_1,\ldots, a_j]\}$ with the help of the generating relation\footnote{For simplicity we sometimes do not indicate dependence on discrete variable $i\in\mathbb{Z}$ in formulae which contain no shifts with respect to this variable.}
\begin{equation}
z^r=a^{[r]}+\sum_{j\geq 1}q_j^{(n, r)}z^{j(n-1)}a^{[r-jn]}.
\label{zr1}
\end{equation}
Clearly, in terms of the wavefunction $\Psi=\{\psi_i\}$ this relation takes the form
\begin{equation}
z^r\psi_i=z^r\psi_{i+r}+\sum_{j\geq 1}z^{r-j}q_j^{(n, r)}(i)\psi_{i+r-jn}.
\label{zr}
\end{equation}
Relation (\ref{zr1}) generate triangular infinite system
\begin{equation}
a_k^{[r]}+\sum_{j=1}^{k-1}a^{[r-jn]}_{k-j}q_j^{(n,r)}+q_k^{(n,r)}=0,\;\;\; k\geq 1.
\label{exact}
\end{equation}
One can check that a more general relation than (\ref{exact}), namely
\begin{equation}
a_k^{[m]}(i)=a_k^{[r]}(i)+\sum_{j=1}^{k-1}a_{k-j}^{[r-jn]}(i)q_j^{(n,r-m)}(i+m)+q_k^{(n,r-m)}(i+m)
\label{iden}
\end{equation}
with any $r, m\in\mathbb{Z}$, is valid \cite{Svinin3}. Resolving the latter in favor of $q_k^{(n,r-p)}(i+p)$ yields
\[
q_k^{(n,r-m)}(i+m)=a_k^{[m]}(i)+\sum_{j=1}^{k-1}q_j^{(n,r-(k-j)n)}(i)a_{k-j}^{[m]}(i)+q_k^{(n,r)}(i).
\]
It should be noted that polynomials $q_j^{(n, r)}$ satisfy following identities:
\[
q_k^{(n, r_1)}(i)+\sum_{j=1}^{k-1}q_j^{(n, r_1)}(i)q_{k-j}^{(n, r_2)}(i+r_1-jn)+q_k^{(n, r_2)}(i+r_1)
\]
\begin{equation}
=q_k^{(n, r_2)}(i)+\sum_{j=1}^{k-1}q_j^{(n, r_2)}(i)q_{k-j}^{(n, r_1)}(i+r_2-jn)+q_k^{(n, r_1)}(i+r_2) \label{rr1rr2}
\end{equation}
for any $r_1, r_2\in\mathbb{Z}$. These identities can be obtained from (\ref{zr}) if we rewrite it in operator form: $z^r\Psi={\mathcal Q}^{(n, r)}(\Psi)$. Then we derive the set of identities (\ref{rr1rr2}) from the relation
\[
z^{r_1+r_2}\Psi={\mathcal Q}^{(n, r_1)}\cdot {\mathcal Q}^{(n, r_2)}(\Psi)={\mathcal Q}^{(n, r_2)}\cdot {\mathcal Q}^{(n, r_1)}(\Psi).
\]
Let us remark that the condition (\ref{eq3.2}) is equivalent to the relation
\[
z^{l-n}a^{[n]}=H^{(l)}+\sum_{k=1}^la_k^{[n]}H^{(l-k)}.
\]

\subsection{The $n$th discrete KP hierarchy}

Let us consider now the restriction of equations (\ref{eq3.1}) on $\mathcal{S}_0^n$ defined by simple relation $z^{1-n}a^{[n]}=h+a_1^{[n]}$ from which we get $h_k=a_k^{[n]}$  first of all. Therefore, we  know that on $\mathcal{S}_0^n$ one has $H^1_k=h_{k+1}=a_{k+1}^{[n]}$. To obtain explicit expressions for all $H_k^s$ as quasi-homogeneous polynomials of $a_k$ we need in theorem \ref{th:2}. As a result we have \cite{Svinin3}
\[
H^s_k=F_k^{(n,s)}[a_1,\ldots, a_{k+s}]\equiv a_{k+s}^{[sn]}+\sum_{j=1}^{s-1}q_j^{(n, sn)}a_{k+s-j}^{[(s-j)n]}.
\]
This totality of relations can be written down as a whole with the help of one generating relation
\begin{equation}
H^{(s)}=F^{(n, s)}=z^{s-sn}\left(a^{[sn]}+\sum_{j=1}^sz^{j(n-1)}q_j^{(n, sn)}a^{[(s-j)n]}\right),
\label{zln}
\end{equation}
where $F^{(n, s)}\equiv z^s+\sum_{j\geq 1}F_j^{(n, s)}z^{-j}$. Thus, the restriction of dynamics given by (\ref{eq3.1}) on $\mathcal{S}_0^n$ leads to evolution equations in the form of differential-difference conservation laws
\begin{equation}
\partial_s\xi_k(i)=F_k^{(n,s)}(i+1)-F_k^{(n,s)}(i)
\label{dKP}
\end{equation}
on infinite number of fields $\{a_k=a_k(i)\}$. The corresponding phase space we denote by $\mathcal{M}$. The hierarchy of the flows on $\mathcal{M}$ given by evolution equations (\ref{dKP}) we call $n$th discrete KP hierarchy. These equations also admit a rich family of reductions.

Following remark is in order. A reformulation of the generating relation (\ref{zln}) in terms of the wavefunction $\Psi=\{\psi_i\}$ is
\begin{equation}
\partial_s\psi_i=z^s\psi_{i+sn}+\sum_{j=1}^sz^{s-j}q_j^{(n, sn)}(i)\psi_{i+(s-j)n}.
\label{auxiliary2}
\end{equation}
Checking compatibility of (\ref{auxiliary2}) with (\ref{zr}) we obtain
\begin{eqnarray}
\partial_sq_k^{(n, r)}(i)&=&q_{k+s}^{(n, r)}(i+sn)+\sum_{j=1}^sq_j^{(n,sn)}(i)q_{k+s-j}^{(n, r)}(i+(s-j)n) \nonumber \\
& &-q_{k+s}^{(n, r)}(i)-\sum_{j=1}^sq_j^{(n,sn)}(i+r-(k+s-j)n)q_{k+s-j}^{(n, r)}(i). \label{dts}
\end{eqnarray}

\subsection{...and its reductions}

Obviously, that restriction of dynamics given by (\ref{eq3.1}) on intersection ${\cal S}_0^n\cap{\cal S}_{l-1}^p$ is equivalent to restriction of $n$th discrete KP hierarchy on some submanifold ${\cal M}_{n, p, l}\subset{\cal M}$. We can easily to write down the equations defining ${\cal M}_{n, p, l}$. They evidently follow from generating relation
\[
z^{l-p}a^{[p]}=F^{(n, l)}+\sum_{j=1}^la_j^{[p]}F^{(n, l-j)}.
\]
From here we obtain that ${\cal M}_{n, p, l}$ is defined by infinite number of equations
\begin{equation}
J_k^{(n, p, l)}[a_1,\ldots, a_{k+l}]=0,\;\;\; \forall k\geq 1
\label{mnpl}
\end{equation}
with
\begin{eqnarray}
J_k^{(n, p, l)}&=&a^{[p]}_{k+l}-F_k^{(n, l)}-\sum_{j=1}^{l-1}a_j^{[p]}F_k^{(n, l-j)}  \nonumber \\
               &=&a^{[p]}_{k+l}(i)-a_{k+l}^{[ln]}(i)-\sum_{j=1}^{l-1}q_j^{(n,ln-p)}(i+p)a_{k+l-j}^{[(l-j)n]}(i). \nonumber
\end{eqnarray}
As a consequence of theorem \ref{th:3}, we have the following.
\begin{thm} \label{th:3}
The following chain of invariant submanifolds inclusions:
\begin{equation}
{\cal M}_{n, p, l}\subset {\cal M}_{n, 2p, 2l}\subset{\cal M}_{n, 3p, 3l}\subset\cdots
\label{eq3.5}
\end{equation}
is valid.
\end{thm}

\subsection{Linear equations on KP wavefunction corresponding to submanifold ${\cal M}_{n, p, l}$}

Some remarks about linear equations on KP formal wave function $\left\{\psi_i\right\}$ which follows as a result of restriction on ${\cal M}_{n, p, l}$ are in order. Let $J^{(n, p, l)}(z)\equiv\sum_{j\geq 1}J_j^{(n, p, l)}z^{-j}$. We observe that an infinite number of equations (\ref{mnpl}) can be presented by single generating relation
\[
\fl
J^{(n, p, l)}(i, z)=z^{l-p}a^{[p]}(i, z)-z^{l-ln}\left(a^{[ln]}(i, z)+\sum_{j=1}^lz^{j(n-1)}q_j^{(n, ln-p)}(i+p)a^{[(l-j)n]}(i, z)\right)=0.
\]
Clearly, in terms of KP wavefunction, we can rewrite the latter relation as
\[
J^{(n, p, l)}(i, z)
=\frac{z^l\psi_{i+p}-z^l\psi_{i+ln}-\sum_{j=1}^lz^{l-j}q_j^{(n, ln-p )}(i+p)\psi_{i+(l-j)n}}{\psi_i}=0.
\]
Thus in terms of KP wave function the restriction of $n$th discrete KP hierarchy on ${\cal M}_{n, p, l}$ is given by the linear equation
\begin{equation}
z^l\psi_{i+ln}+\sum_{j=1}^lz^{l-j}q_j^{(n, ln-p )}(i+p)\psi_{i+(l-j)n}=z^l\psi_{i+p}.
\label{auxiliary1}
\end{equation}
Second linear equation which we should have in mind is (\ref{auxiliary2}).

When considering restriction of $n$th discrete KP hierarchy on ${\cal M}_{n, p, l}$ one can find many examples of integrable lattices known from the literature (some examples can be found in the paper \cite{Svinin3}) and construct ``new'' ones. Classical examples are Volterra and Toda lattices. In what follows, we restrict ourselves by consideration only a class of one-field lattices corresponding to ${\cal M}_{n, p, 1}$.

\section{The polynomials $S^l_s$ and $T^l_s$} \label{sec:4}

In the next section we consider restriction of $n$th discrete KP hierarchy on the submanifold ${\cal M}_{n, p, 1}$, with $n\geq 1$ and $p>n$. For further use, let us prepare in this section two classes of polynomials through\footnote{We use perhaps unusual but quite convenient notation writing ``first'' and ``last'' argument through the vertical bar.}
\begin{equation}
S^l_s(y_1|y_{sp-(s-l+1)n+1})=\sum_{1\leq\lambda_l\leq\cdots\leq\lambda_1\leq s+1}\left(\prod_{j=1}^ly_{(p-n)\lambda_j+jn-p+1}\right)
\label{S}
\end{equation}
and
\begin{equation}
T^l_s(y_1|y_{sp-(s-l+1)n+1})=\sum_{1\leq\lambda_1<\cdots<\lambda_l\leq s+1}\left(\prod_{j=1}^ly_{(p-n)\lambda_j+jn-p+1}\right)
\label{T}
\end{equation}
for $s\geq 0$.\footnote{It is convenient to think that $T^l_s\equiv 0$ for $s=0,\ldots, l-2$.}
Let us show some identities for these polynomials. Firstly consider $S^l_s$. A partition of the set
\[
B_{l, s}=\left\{\lambda_j\; :\; 1\leq\lambda_l\leq\cdots\leq\lambda_1\leq s+1\right\}
\]
into a pair of disjoint subsets $B_{l, s}=B_1^l\cup B_{l, s-1}$ with
\[
B_1^l=\left\{\lambda_j\; :\; \lambda_1=s+1,\;1\leq\lambda_l\leq\cdots\leq\lambda_2\leq s+1\right\}
\]
and
\[
B_{l, s-1}=\left\{\lambda_j\; :\; 1\leq\lambda_l\leq\cdots\leq\lambda_1\leq s\right\},
\]
as can be checked,  leads to the identity
\begin{equation}
\fl
S^l_s(y_1|y_{sp-(s-l+1)n+1})=S^l_{s-1}(y_1|y_{(s-1)p-(s-l)n+1})+y_{(p-n)s+1}S^{l-1}_s(y_{n+1}|y_{sp-(s-l+1)n+1}).
\label{id3}
\end{equation}
Let
\[
B_k^l=\left\{\lambda_j\; :\; \lambda_1=s-k+2,\;1\leq\lambda_l\leq\cdots\leq\lambda_2\leq s-k+2\right\}
\]
and
\[
B_{l, s-k+1}=\left\{\lambda_j\; :\; 1\leq\lambda_l\leq\cdots\leq\lambda_1\leq s-k+2\right\}
\]
for $k=1,\ldots, s+1$. Clearly, $B_{l, s-k+1}=B_k^l\cup B_{l, s-k}$ and $B_{l, 0}=B_{s+1}^l$. Thus, we have the following:
\[
B_{l, s}=\bigcup_{j=1}^{s+1}B_j^l.
\]
This decomposition of $B_{l, s}$ in turn yields the identity
\begin{equation}
\fl
S^l_s(y_1|y_{sp-(s-l+1)n+1})=\sum_{j=1}^{s+1}y_{(s-j+1)(p-n)+1}S^{l-1}_{s-j+1}(y_{n+1}|y_{(s-j+1)p-(s-l-j+2)n+1}).
\label{id1}
\end{equation}
On the other hand a partition of $B_{l, s}$ into
\[
\bar{B}^l_1=\left\{\lambda_j : \lambda_l=1,\;1\leq\lambda_{l-1}\leq\cdots\leq\lambda_1\leq s+1\right\}
\]
and
\[
\bar{B}_{l, s-1}=\left\{\lambda_j : 2\leq\lambda_l\leq\cdots\leq\lambda_1\leq s+1\right\}
\]
gives the following identity:
\begin{equation}
\fl
S^l_s(y_1|y_{sp-(s-l+1)n+1})=S^l_{s-1}(y_{p-n+1}|y_{sp-(s-l+1)n+1})
+y_{(l-1)n+1}S^{l-1}_s(y_1|y_{sp-(s-l+2)n+1}).
\label{id4}
\end{equation}
Making  use of the partition $\bar{B}_{l, s-k+1}=\bar{B}_k^l\cup \bar{B}_{l, s-k}$ and $\bar{B}_{l, 0}=\bar{B}_{s+1}^l$ with
\[
\bar{B}^l_k=\left\{\lambda_j : \lambda_l=k,\;
k\leq\lambda_{l-1}\leq\cdots\leq\lambda_1\leq s+1\right\}
\]
and
\[
\bar{B}_{l, s-k+1}=\left\{\lambda_j : k\leq\lambda_l\leq\cdots\leq\lambda_1\leq s+1\right\}
\]
we are led to the partition
\[
B_{l, s}=\bigcup_{j=1}^{s+1}\bar{B}_j^l
\]
and corresponding identity
\begin{equation}
\fl
S^l_s(y_1|y_{sp-(s-l+1)n+1})
=\sum_{j=1}^{s+1}y_{(j-1)p-(j-l)n+1}S^{l-1}_{s-j+1}(y_{(j-1)(p-n)+1}|y_{sp-(s-l+2)n+1}).
\label{id2}
\end{equation}
It is worth to remark that identities (\ref{id1}) and (\ref{id2}) being in nature recurrence relations both uniquely define polynomials $S^l_s$ starting from $S^0_s\equiv 1$.

Consider now polynomials $T^l_s$. A partition of the  set
\[
D_{l, s}=\left\{\lambda_j\; :\; 1\leq\lambda_1<\cdots<\lambda_l\leq s+1\right\},
\]
into two subsets
\[
D_1^l=\left\{\lambda_j\; :\; \lambda_1=1,\;2\leq\lambda_2<\cdots<\lambda_l\leq s+1\right\}
\]
and
\[
D_{l, s-1}=\left\{\lambda_j\; :\; 2\leq\lambda_1<\cdots<\lambda_l\leq s+1\right\}
\]
leads to the identity
\begin{equation}
\fl
T^l_s(y_1|y_{sp-(s-l+1)n+1})=T^l_{s-1}(y_{p-n+1}|y_{sp-(s-l+1)n+1})
+y_1T^{l-1}_{s-1}(y_{p+1}|y_{sp-(s-l+1)n+1}).
\label{id5}
\end{equation}
Let
\[
D_k^l=\left\{\lambda_j\; :\; \lambda_1=k,\;k+1\leq\lambda_2<\cdots<\lambda_l\leq s+1\right\}
\]
and
\[
D_{l, s-k+1}=\left\{\lambda_j\; :\; k\leq\lambda_1<\cdots<\lambda_l\leq s+1\right\}
\]
for $k=1,\ldots, s-l+2$. Taking into account that $D_{l, s-k+1}=D^l_k\cup D_{l, s-k}$ and $D_{l, l-1}=D_{s-l+2}^l$ we are led to the partition
\[
D_{l, s}=\bigcup_{j=1}^{s-l+2}D_j^l
\]
and corresponding identity
\begin{equation}
\fl
T^l_s(y_1|y_{sp-(s-l+1)n+1})=\sum_{j=1}^{s-l+2}y_{(j-1)(p-n)+1}T^{l-1}_{s-j}(y_{jp-(j-1)n+1}|y_{sp-(s-l+1)n+1})
\label{id8}
\end{equation}
Finally, consider the partition $D_{l, s}=\bar{D}^l_1\cup\bar{D}_{l, s-1}$ with
\[
\bar{D}_1^l=\left\{\lambda_j\; :\; \lambda_l=s+1,\;1\leq\lambda_1<\cdots<\lambda_{l-1}\leq s\right\}
\]
and
\[
\bar{D}_{l, s-1}=\left\{\lambda_j\; :\; 1\leq\lambda_1<\cdots<\lambda_l\leq s\right\}
\]
and corresponding identity
\begin{equation}
\fl
T^l_s(y_1|y_{sp-(s-l+1)n+1})=T^l_{s-1}(y_1|y_{(s-1)p-(s-l)n+1})+y_{sp-(s-l+1)n+1}T^{l-1}_{s-1}(y_1|y_{(s-1)p-(s-l+1)n+1}).
\label{id6}
\end{equation}
Let
\[
\bar{D}_k^l=\left\{\lambda_j\; :\; \lambda_l=s-k+2,\;1\leq\lambda_1<\cdots<\lambda_{l-1}\leq s-k+1\right\}
\]
and
\[
\bar{D}_{l, s-k+1}=\left\{\lambda_j\; :\; 1\leq\lambda_1<\cdots<\lambda_l\leq s-k+1\right\}.
\]
We have $\bar{D}_{l, s-k+1}=\bar{D}^l_k\cup \bar{D}_{l, s-k}$ and $\bar{D}_{l, l-1}=\bar{D}_{s-l+2}^l$. The partition
\[
D_{l, s}=\bigcup_{j=1}^{s-l+2}\bar{D}_j^l
\]
leads to the identity
\begin{equation}
\fl
T^l_s(y_1|y_{sp-(s-l+1)n+1})=\sum_{j=1}^{s-l+2}y_{(s-j+1)p-(s-l-j+2)n+1}T^{l-1}_{s-j}(y_1|y_{(s-j)p-(s-l-j+2)n+1}).
\label{id7}
\end{equation}

Let us identify $y_k=r_{i+k-1}$ for $k=1,\ldots, sp-(s-l+1)n+1$, where $r=r(i)\equiv r_i$ is some unknown function of discrete variable $i\in\mathbb{Z}$. We define discrete polynomial functions $S^l_s[r]$ and $T^l_s[r]$ by
\[
S^l_s(i)\equiv S^l_s(r_i|r_{i+sp-(s-l+1)n})\;\;\;\mbox{and}\;\;\; T^l_s(i)\equiv T^l_s(r_i|r_{i+sp-(s-l+1)n}),
\]
respectively. Let us write down below all identities for $S^l_s[r]$ and $T^l_s[r]$ corresponding to relations (\ref{id3})-(\ref{id7}) in their order. We have the following:
\begin{eqnarray}
S^l_s(i)&=&S^l_{s-1}(i)+r_{i+s(p-n)}S^{l-1}_s(i+n) \nonumber \\
        &=&\sum_{j=1}^{s+1}r_{i+(s-j+1)(p-n)}S^{l-1}_{s-j+1}(i+n) \nonumber \\
        &=&S^l_{s-1}(i+p-n)+r_{i+(l-1)n}S^{l-1}_s(i) \nonumber \\
        &=&\sum_{j=1}^{s+1}r_{i+(j-1)p-(j-l)n}S^{l-1}_{s-j+1}(i+(j-1)(p-n)) \nonumber
\end{eqnarray}
and
\begin{eqnarray}
T^l_s(i)&=&T^l_{s-1}(i+p-n)+r_iT^{l-1}_{s-1}(i+p) \nonumber \\
        &=&\sum_{j=1}^{s-l+2}r_{i+(j-1)(p-n)}T^{l-1}_{s-j}(i+jp-(j-1)n) \nonumber \\
        &=&T^l_{s-1}(i)+r_{i+sp-(s-l+1)n}T^{l-1}_{s-1}(i) \nonumber \\
        &=&\sum_{j=1}^{s-l+2}r_{i+(s-j+1)p-(s-l-j+2)n}T^{l-1}_{s-j}(i). \nonumber
\end{eqnarray}
Since $T^l_s\equiv 0$ for $s=0,\ldots,l-2$, then we also can write
\begin{eqnarray}
T^l_s(i)&=&\sum_{j=1}^sr_{i+(j-1)(p-n)}T^{l-1}_{s-j}(i+jp-(j-1)n) \nonumber \\
        &=&\sum_{j=1}^sr_{i+(s-j+1)p-(s-l-j+2)n}T^{l-1}_{s-j}(i). \label{kto}
\end{eqnarray}

\section{The manifold $\mathcal{M}_{n, p, 1}$} \label{sec:5}

\subsection{Technical lemmas}

Our aim is to show how polynomials $S^l_s[r]$ and $T^l_s[r]$ constructed in previous section appear when constructing integrable hierarchies for some differential-difference equations in its explicit form.

Let us consider the submanifold $\mathcal{M}_{n, p, 1}\subset\mathcal{M}$ with $n\geq 1$ and $p>n$ defined by equations
$J_k^{(n, p, 1)}=a_{k+1}^{[p]}-a_{k+1}^{[n]}\equiv 0$ for $k\geq 1$ which, using (\ref{r1r2}) can be rewritten equivalently as
\[
\sum_{j=1}^ka_{k-j+1}^{[n]}(i)a_j^{[p-n]}(i+n)+a_{k+1}^{[p-n]}(i+n)=0.
\]
Making use again of the identity (\ref{r1r2}) we see that solution of this equation is given by
\begin{equation}
a_k^{[p-n]}(i)=a_{k-1}^{[-n]}(i)r_i\;\;\;\mbox{with}\;\;\; r_i\equiv a_1^{[p-n]}(i)
\label{solution}
\end{equation}
for all $k\geq 2$.
\begin{lem}
On $\mathcal{M}_{n, p, 1}$ 
\begin{equation}
a_k^{[(p-n)s]}(i)=\sum_{j=1}^sa_{k-1}^{[-jn+(j-1)p]}(i)r_{i+(j-1)(p-n)},
\label{1}
\end{equation}
and
\begin{equation}
a_k^{[(n-p)s]}(i)=-\sum_{j=1}^{s}a_{k-1}^{[-jp+(j-1)n]}(i)r_{i+j(n-p)}
\label{2}
\end{equation}
for all integers $s\geq 1$.
\end{lem}

\textbf{Proof}.
In virtue of (\ref{r1r2}) and (\ref{solution}), 
\[
\fl
a_k^{[(p-n)s]}(i)=a_k^{[(p-n)(s-1)]}(i)+\sum_{j=1}^{k-1}a_{k-j}^{[(p-n)(s-1)]}(i)a_j^{[p-n]}(i+(p-n)(s-1))
\]
\[
+a_k^{[p-n]}(i+(p-n)(s-1)) =a_k^{[(p-n)(s-1)]}(i)+\left\{a_{k-1}^{[(p-n)(s-1)]}(i) \right.
\]
\[
\fl
\left. +\sum_{j=1}^{k-2}a_{k-j-1}^{[(p-n)(s-1)]}(i)a_j^{[-n]}(i+(p-n)(s-1))+a_{k-1}^{[-n]}(i+(p-n)(s-1)) \right\}r_{i+(p-n)(s-1)}
\]
\[
=a_k^{[(p-n)(s-1)]}(i)+a_{k-1}^{[(s-1)p-sn]}(i)r_{i+(p-n)(s-1)}.
\]
Making use of this recurrence relation we immediately obtain (\ref{1}) and (\ref{2}).

\begin{lem} \label{lem:2}
On $\mathcal{M}_{n, p, 1}$ 
\begin{equation}
q_k^{(n,(p-n)s)}(i)=(-1)^kS_{s-1}^k(i-(k-1)n),
\label{pn}
\end{equation}
\begin{equation}
q_k^{(n, (n-p)s)}(i)=T_{s-1}^k(i-(k-1)n+(n-p)s)
\label{np}
\end{equation}
for $s\geq 1$.
\end{lem}

Let us remark that this lemma, in particular, says  that on $\mathcal{M}_{n, p, 1}$ we have $q_k^{(n, (n-p)s)}\equiv 0$ for $s=1,\ldots, k-1$.

\textbf{Proof of lemma 2}. In particular case $r=1$  identity (\ref{iden}) is specified as
\[
a_k^{[m]}(i)=\sum_{j=1}^{k-1}a_{k-j}^{[-jn]}(i)q_j^{(n,-m)}(i+m)+q_k^{(n,-m)}(i+m).
\]
Let $m=(p-n)s$. We observe that in virtue of the latter identity, (\ref{pn}) is equivalent to the relation
\begin{equation}
a_k^{[(p-n)s]}(i)=\sum_{j=1}^{k-1}a_{k-j}^{[-jn]}(i)T^{k-j}_{s-1}(i-(j-1)n)+T^k_{s-1}(i-(k-1)n).
\label{ak}
\end{equation}
So, let us prove (\ref{ak}) instead of (\ref{np}). For $k=1$, the latter becomes $a_1^{[(p-n)s]}(i)=T^1_{s-1}(i)$ what is evident, because\footnote{Here $a_i\equiv a_1(i)$.}
\begin{equation}
a_1^{[(p-n)s]}(i)=\sum_{j=1}^{(p-n)s}a_{i+j-1}=\sum_{j=1}^sr_{i+(j-1)(p-n)}\equiv T^1_{s-1}(i)
\label{ak1}
\end{equation}
simply by definition and without reference to any invariant submanifold. Further, we proceed by induction. Suppose we have proved (\ref{ak}) for $k=1,\ldots, k_0$, where $k_0\geq 1$. Then it is easy to see, making use of (\ref{r1r2}), that for these values of $k$ the relations of the form
\begin{eqnarray}
\fl
a_k^{[m+(p-n)s]}(i)-a_k^{[m]}(i)&=&T^k_{s-1}(i+m-(k-1)n) \nonumber \\
\fl
                                &&+\sum_{j=1}^{k-1}a_j^{[m-(k-j)n]}(i)T^{k-j}_{s-1}(i+m-(k-j-1)n) \nonumber
\end{eqnarray}
hold for any $m\in\mathbb{Z}$. In particular, let $m=-n$; then
\begin{eqnarray}
\fl
a_k^{[-n+(p-n)s]}(i)-a_k^{[-n]}(i)&=&T^k_{s-1}(i-kn) \nonumber \\
\fl
                                  & &+\sum_{j=1}^{k-1}a_j^{[-(k-j+1)n]}(i)T^{k-j}_{s-1}(i-(k-j)n). \label{yu}
\end{eqnarray}
With the help of (\ref{kto}), (\ref{1}), (\ref{ak1}) and (\ref{yu}) we have the following:
\[
\fl
a_{k+1}^{[(p-n)s]}(i)-a_k^{[-n]}(i)T^1_{s-1}(i)=\sum_{j=1}^{s-1}\left\{a_k^{[-(j+1)n+jp]}(i)-a_k^{[-n]}(i)\right\}r_{i+j(p-n)}
\]
\[
\fl
=\sum_{j=1}^{s-1}\left\{T^k_{j-1}(i-kn)+\sum_{j_1=1}^{k-1}a_{j_1}^{[-(k-j_1+1)n]}(i)T^{k-j_1}_{j-1}(i-(k-j_1)n)\right\}r_{i+j(p-n)}
\]
\[
\fl
=\sum_{j=1}^{s-1}T^k_{j-1}(i-kn)r_{i+j(p-n)}+\sum_{j_1=1}^{k-1}a_{j_1}^{[-(k-j_1+1)n]}(i)\left(\sum_{j=1}^{s-1}T^{k-j_1}_{j-1}(i-(k-j_1)n)r_{i+j(p-n)}\right)
\]
\[
\fl
=T^{k+1}_{s-1}(i-kn)+\sum_{j_1=1}^{k-1}a_{j_1}^{[-(k-j_1+1)n]}(i)T_{s-1}^{k-j_1+1}(i-(k-j_1)n).
\]
Thus, we have proved that if (\ref{ak}) is valid for $k=1,\ldots, k_0$ then it is valid for $k=k_0+1$ and therefore we can use now induction with respect to $k$. The relation (\ref{pn}) is proved by using similar reasonings.

\subsection{Additional identities for $S^l_s[r]$ and $T^l_s[r]$}

Making  use of lemma \ref{lem:2} and identities (\ref{rr1rr2}) with $r_1=(p-n)(s_1+1)$ and $r_2=(p-n)(s_2+1)$ we are able to obtain following identities:
\[
S^l_{s_1}(i)+\sum_{j=1}^{l-1}(-1)^jS^{l-j}_{s_1}(i)T^j_{s_2}(i+(l-j)n)+(-1)^lT^l_{s_2}(i)
\]
\[
\fl
=S^l_{s_1}(i+(s_2+1)(p-n))+\sum_{j=1}^{l-1}(-1)^jS^{l-j}_{s_1}(i+(s_2+1)(p-n)+jn)T^j_{s_2}(i+(s_1+1)(p-n))
\]
\begin{equation}
+(-1)^lT^l_{s_2}(i+(s_1+1)(p-n)).
\label{form}
\end{equation}
In particular, let $s_1=s_2=s$. Then we have
\[
S^l_{s}(i)+\sum_{j=1}^{l-1}(-1)^jS^{l-j}_{s}(i)T^j_{s}(i+(l-j)n)+(-1)^lT^l_{s}(i)=0
\]
and
\[
S^l_{s}(i)+\sum_{j=1}^{l-1}(-1)^jS^{l-j}_{s}(i+jn)T^j_{s}(i)+(-1)^lT^l_{s}(i)=0.
\]

\subsection{Linear equations on KP wave function and its compatibility}

Let us discuss linear equations on KP wave function $\Psi=\{\psi_i\}$. On $\mathcal{M}_{n, p, 1}$ we have the linear equation
\begin{equation}
z\psi_{i+n}+T^1_0(i+n)\psi_i=z\psi_{i+p}
\label{linear}
\end{equation}
with $T^1_0(i)=r_i$ which is a specification of (\ref{auxiliary1}). On the other hand, from theorem \ref{th:3} we have
\[
\mathcal{M}_{n, p, 1}\subset\mathcal{M}_{n, 2p, 2}\subset\mathcal{M}_{n, 3p, 3}\subset\cdots
\]
and corresponding infinite set of linear equations
\begin{equation}
z^k\psi_{i+kn}+\sum_{j=1}^kT^j_{k-1}(i+(k-j+1)n)z^{k-j}\psi_{i+(k-j)n}=z^k\psi_{i+kp}
\label{linear1}
\end{equation}
for $k\geq 2$ which can be obtained also as a consequences of linear equation (\ref{linear}). Remark that we obtain coefficients of (\ref{linear1}) making use of lemma (\ref{lem:2}). Let us remember that $T^l_s[r]$'s in  (\ref{linear1}) are discrete polynomials defined for some fixed $n\geq 1$  and $p>n$ via (\ref{T}).
Let us consider linear evolution equation
\[
\psi_i^{\prime}=z\psi_{i+n}-s_i\psi_i\;\;\;\mbox{with}\;\;\; s_i\equiv a_1^{[n]}(i)
\]
for the first flow of $n$th discrete KP hierarchy. By straightforward calculations we check that the compatibility of the latter equation with (\ref{linear1}) is guaranteed when the relation
\[
\partial T_k^l(i)=T^{l+1}_k(i)-T^{l+1}_k(i-n)+T_k^l(i)\left(s_{i-n}-s_{i+(k+1)p-(k-l+2)n}\right)
\]
holds. In particular,
\[
\partial T_1^0(i)=\partial r_i=r_i\left(s_{i-n}-s_{i+p-n}\right)
\]
or
\[
\sum_{j=1}^{p-n}a_{i+j-1}^{\prime}=\sum_{j=1}^{p-n}a_{i+j-1}\left(\sum_{j=1}^na_{i-j}-\sum_{j=p-n}^{p-1}a_{i+j}\right).
\]

\subsection{Integrable hierarchies of differential-difference equations associated with $\mathcal{M}_{n, p, 1}$}

In this subsection we  are going to show a class of integrable hierarchies corresponding to $\mathcal{M}_{n, p, 1}$. Making use of lemma 2 we obtain
\begin{eqnarray}
\fl
\partial^{*}_s\psi_i &\equiv & \partial_{(p-n)s}\psi_i \nonumber \\
\fl
                     &=      &z^{(p-n)s}\psi_{i+(p-n)sn}+\sum_{j=1}^{(p-n)s}(-1)^jz^{(p-n)s-j}S^j_{sn-1}(i-(j-1)n)\psi_{i+(p-n)sn-jn}.
\label{ev}
\end{eqnarray}
The condition of compatibility of (\ref{ev}) with (\ref{linear}) yields
\begin{eqnarray}
\fl
\partial^{*}_sT^1_0(i)&=&\partial^{*}_sr_i \nonumber \\
\fl
                       &=&(-1)^{(p-n)s}r_i\left\{S_{sn-1}^{(p-n)s}(i-(p-n)sn+p)-S_{sn-1}^{(p-n)s}(i-(p-n)sn)\right\}
\label{ev1}
\end{eqnarray}
while when checking  compatibility of (\ref{ev}) with (\ref{linear1}) we get
\[
\fl
\partial^{*}_sT^l_k(i)=T^{l+(p-n)s}_k(i)
\]
\[
\fl
+\sum_{j=1}^{(p-n)s}(-1)^jS_{sn-1}^j(i+(p-n)(k+1)+(l-j)n)T^{l+(p-n)s-j}_k(i)-T^{l+(p-n)s}_k(i-(p-n)sn)
\]
\begin{equation}
\fl
-\sum_{j=1}^{(p-n)s}(-1)^jS_{sn-1}^j(i-(p-n)sn)T^{l+(p-n)s-j}_k(i-(p-n)sn+jn).
\label{pni}
\end{equation}
Remark that all other relations obtained as a result of checking compatibility  of (\ref{ev}) with (\ref{linear1}) are just identities of the form (\ref{form}). Remark also that  equations (\ref{pni}) could be obtained in an easier way by using (\ref{dts}) with lemma \ref{lem:2}. Following along this line, in addition to (\ref{pni}), we get the following
\begin{eqnarray}
\fl
\partial^{*}_sS^l_k(i)&=&(-1)^{(p-n)s}\left\{S^{l+(p-n)s}_k(i)+\sum_{j=1}^{(p-n)s}S_{sn-1}^j(i+(k-j)n)S^{l+(p-n)s-j}_k(i) \right. \nonumber \\
\fl                      & &   -S^{l+(p-n)s}_k(i-(p-n)sn)  \nonumber \\
\fl
                      & &\left. -\sum_{j=1}^{(p-n)s}S_{sn-1}^j(i+(p-n)(k-sn+1))S^{l+(p-n)s-j}_k(i-(p-n)sn+jn)\right\}. \label{ev5}
\end{eqnarray}

\subsection{The case $\mathcal{M}_{n, n+1, 1}$} \label{sec:7}

Let us consider the submanifold $\mathcal{M}_{n, n+1, 1}$. Since in this case $p-n=1$ then $r_i=a_i$. Linear equation (\ref{linear}) in this case becomes
\begin{equation}
z\psi_{i+n}+a_{i+n}\psi_i=z\psi_{i+n+1}
\label{lineaar}
\end{equation}
while evolution equation (\ref{ev}) takes the form
\[
\partial_s\psi_i=z^s\psi_{i+sn}+\sum_{j=1}^sz^{s-j}(-1)^jS^j_{sn-1}(i-(j-1)n)\psi_{i+(s-j)n}
\]
and (\ref{ev1}) is specified as
\begin{equation}
\partial_sa_i=(-1)^sa_i\left\{S_{sn-1}^s(i-(s-1)n+1)-S_{sn-1}^s(i-sn)\right\}.
\label{higher}
\end{equation}
The first flow in this hierarchy is given by differential-difference equation
\begin{eqnarray}
a_i^{\prime}&=&-a_i\left(S_{n-1}^1(i+1)-S_{n-1}^1(i-n)\right) \nonumber \\
        &=&a_i\left(\sum_{j=1}^na_{i-j}-\sum_{j=1}^na_{i+j}\right). \label{Bl}
\end{eqnarray}
This equation with quadratic nonlinearity is nothing but INB lattice mentioned in the introduction. It is known by \cite{Bogoyavlenskii}  that INB lattice, being in a sense integrable generalization of the Volterra lattice $a_i^{\prime}=a_i(a_{i-1}-a_{i+1})$, for any $n\geq 1$, gives integrable discretization of the Korteweg-de Vries (KdV) equation.

Therefore as a particular case we constructed integrable hierarchy for INB equation (\ref{Bl}) in its explicit form \cite{Svinin4}.  Remark that polynomials $S^l_s$ and $T^l_s$ in this case are specified as
\[
S^l_s(y_1|y_{s+(l-1)n+1})=\sum_{1\leq\lambda_l\leq\cdots\leq\lambda_1\leq s+1}\left(\prod_{j=1}^ly_{\lambda_j+n(j-1)}\right)
\]
and
\[
T^l_s(y_1|y_{s+(l-1)n+1})=\sum_{1\leq\lambda_1<\cdots<\lambda_l\leq s+1}\left(\prod_{j=1}^ly_{\lambda_j+n(j-1)}\right).
\]
These polynomials  were introduced in \cite{Svinin4}.

As  consequences of linear equation (\ref{lineaar}) corresponding to the chain of inclusions
\[
\mathcal{M}_{n, n+1, 1}\subset\mathcal{M}_{n, 2n+2, 2}\subset\mathcal{M}_{n, 3n+3, 3}\subset\cdots
\]
we have an infinite number of linear equations
\[
z^k\psi_{i+kn}+\sum_{j=1}^kT^j_{k-1}(i+(k-j+1)n)z^{k-j}\psi_{i+(k-j)n}=z^k\psi_{i+kn+k}
\]
for $k\geq 2$.
Finally, (\ref{pni}) and (\ref{ev5}) in this case becomes
\begin{eqnarray}
\partial_s T_k^l(i)&=&T_k^{l+s}(i)+\sum_{j=1}^s(-1)^jS^j_{sn-1}(i+(l-j)n+k+1)T_k^{l+s-j}(i) \nonumber \\
                   & &-T_k^{l+s}(i-sn)-\sum_{j=1}^s(-1)^jS^j_{sn-1}(i-sn)T_k^{l+s-j}(i-(s-j)n) \nonumber
\end{eqnarray}
and
\begin{eqnarray}
\partial_sS^l_k(i)&=&(-1)^s\left\{S^{l+s}_k(i)+\sum_{j=1}^sS_{sn-1}^j(i+(k-j)n)S^{l+s-j}_k(i) \right. \nonumber \\
 & &\left.-S^{l+s}_k(i-sn)-\sum_{j=1}^sS_{sn-1}^j(i+k-sn+1)S^{l+s-j}_k(i-(s-j)n)\right\}, \nonumber
\end{eqnarray}
respectively. In the paper \cite{Svinin4} we have proved the following.

\begin{thm}
Each one of the constraints
\[
T^{l+1}_s(i+1)=T^{l+1}_s(i),\;\;\; s\geq l
\]
and
\[
S^{l+1}_s(i+1)=S^{l+1}_s(i),\;\;\; s\geq 0
\]
is consistent with the INB lattice hierarchy (\ref{higher}).
\end{thm}
This theorem gives an infinite number of constraints for INB lattice hierarchy including periodicity and stationary conditions.

\subsection{The case $\mathcal{M}_{1, g+1, 1}$} \label{sec:8}

As a second example, let us consider the submanifold $\mathcal{M}_{1, g+1, 1}$. In this case $p-n=g$ and consequently $r_i=a_1^{[g]}(i)$. Linear equation (\ref{linear}) and its consequences (\ref{linear1}) are specified as
\[
z\psi_{i+1}+r_{i+1}\psi_i=z\psi_{i+g+1}
\]
and
\[
z^k\psi_{i+k}+\sum_{j=1}^kT^j_{k-1}(i+k-j+1)z^{k-j}\psi_{i+k-j}=z^k\psi_{i+kg+k},
\]
respectively. Linear evolution  equation (\ref{ev}) becomes
\[
\partial^{*}_s\psi_i \equiv \partial_{gs}\psi_i=z^{gs}\psi_{i+gs}+\sum_{j=1}^{gs}(-1)^jz^{gs-j}S^j_{s-1}(i-j+1)\psi_{i+gs-j}.
\]
Corresponding hierarchy of differential-difference equations is given by
\[
\partial^{*}_sr_i=(-1)^{gs}r_i\left\{S_{s-1}^{gs}(i-gs+g+1)-S_{s-1}^{gs}(i-gs)\right\}.
\]
In particular, the first flow of this hierarchy is managed by
\begin{eqnarray}
\partial^{*}_1r_i&=&(-1)^gr_i\left\{S_0^g(i+1)-S_0^g(i-g)\right\} \nonumber \\
              &=&(-1)^gr_i\left(\prod_{j=1}^gr_{i+j}-\prod_{j=1}^gr_{i-j}\right). \nonumber
\end{eqnarray}
This equation is known by  \cite{Bogoyavlenskii}. It is, as well as the INB equation, represents, for any $g\geq 1$, integrable discretization of the KdV equation.  It is a simple exercise to check that in virtue of the latter (cf. \cite{Bogoyavlenskii})
\[
\partial^{*}_1S^g_0(i)=(-1)^gS^g_0(i)\left(\sum_{j=1}^gS^g_0(i+j)-\sum_{j=1}^gS^g_0(i-j)\right),
\]
i.e., $S^g_0$ satisfies INB equation. Finally, let us write down below (\ref{pni}) and (\ref{ev5}) in this case. We have
\begin{eqnarray}
\partial^{*}_sT^l_k(i)&=&T^{l+gs}_k(i) \nonumber \\
&&+\sum_{j=1}^{gs}(-1)^jS_{s-1}^j(i+g(k+1)+l-j)T^{l+gs-j}_k(i)-T^{l+gs}_k(i-gs) \nonumber \\
&&-\sum_{j=1}^{gs}(-1)^jS_{s-1}^j(i-gs)T^{l+gs-j}_k(i-gs+j) \nonumber
\end{eqnarray}
and
\begin{eqnarray}
\partial^{*}_sS^l_k(i)&=&(-1)^{gs}\left\{S^{l+gs}_k(i)+\sum_{j=1}^{gs}S_{s-1}^j(i+k-j)S^{l+gs-j}_k(i) \right. \nonumber \\
 & &\left. -S^{l+gs}_k(i-gs)-\sum_{j=1}^{gs}S_{s-1}^j(i+g(k-s+1))S^{l+gs-j}_k(i-gs+j)\right\}. \nonumber
\end{eqnarray}

\section{Conclusion}

Our main goal in the paper was to construct explicit form for integrable hierarchies for some class of differential-difference equations. Perhaps a most important case in this class of equations is given by Volterra lattice. What one learned from this presentation is that evolution equations of integrable hierarchies from this class are essentially formulated with the help of special homogeneous polynomials which we present by explicit formulas (\ref{S}) and (\ref{T}). In forthcoming papers, we will show how this information can be exploited for investigation of some problems concerning equations from this class. Remark that in the present paper we consider only one-component lattices. It might be interesting to extend these results to multi-component ones.

\section*{Acknowledgments}

This work has been supported by Russian Foundation for Basic Research  grant No. 09-01-00192-a.

\end{document}